\documentclass[%
 reprint,
 amsmath,amssymb,
 aps,
prl,
]{revtex4-2}

\usepackage{graphicx}
\usepackage{dcolumn}
\usepackage{bm}

\begin{document}

\preprint{APS/123-QED}

\title{Observation of Photonic Chiral Flatbands}

\author{Minho Choi}
 \email{kernel@uw.edu}
 \affiliation{%
 Department of Electrical and Computer Engineering, University of Washington, Seattle, WA 98195, USA
}%

\author{Andrea Al\`{u}}
 \email{aalu@gc.cuny.edu}
 \affiliation{%
Photonics Initiative, Advanced Science Research Center, City University of New York, New York, NY 10031, USA}%
 \affiliation{%
Physics Program, Graduate Center, City University of New York, New York, NY 10031, USA}%

\author{Arka Majumdar}
 \email{arka@uw.edu}
 \affiliation{%
 Department of Electrical and Computer Engineering, University of Washington, Seattle, WA 98195, USA
}%
\affiliation{%
 Department of Physics, University of Washington, Seattle, WA 98195, USA
}%

\date{\today}

\begin{abstract}
Distinct selectivity to the spin angular momenta of photons have garnered significant attention in recent years, for their relevance in basic science and for imaging and sensing applications. While nonlocal metasurfaces with strong chiral responses to the incident light have been reported, these responses are typically limited to a narrow range of incident angles. In this study, we demonstrate a nonlocal metasurface that showcases strong chirality, characterized by circular dichroism larger than 0.6, over a wide range of incident angles $\pm5^o$. Its quality factor, circular dichroism and resonant frequency can be optimized by design. These findings pave the way to further advance the development of valley-selective optical cavities and augmented reality applications.
\end{abstract}

\maketitle



Chirality, stemming from broken mirror or rotational symmetry, is an important property in many scientific and technological fields \cite{liu2023detection,schilthuizen2005convoluted, cook1903spirals}. Optical chirality is associated with circular polarization selectivity, i.e., to the spin angular momentum of a photon, and it is quantified by circular dichroism ($CD$), i.e., absorption or reflection/transmission selectivity to right-handed ($RCP$) and left-handed circular polarization ($LCP$) \cite{tang2010optical,chen2023observation}.

In the past, optical chirality has been generally obtained using macroscopic optical materials \cite{glover1985comparison}. Recently, photonic structures with sub-wavelength thickness, also known as metasurfaces, have gained popularity to control light-matter interactions in compact form factors. In the far-field, chiral metasurfaces have different responses in reflection (or transmission) to $RCP$ and $LCP$ excitations. Thus, in the far field we define $CD$ of the metasurface by its degree of circular polarization in reflection \cite{chen2023observation,zhang2022chiral}:
\begin{equation}
{
 CD=\frac{I_{RCP}-I_{LCP}}{I_{RCP}+I_{LCP}}
},
\label{equation1}
\end{equation}
where $I_{RCP}$ and $I_{LCP}$ are the reflection intensity for $RCP$ and $LCP$ input light, respectively. We can also evaluate the near-field chirality of the metasurface by evaluating its optical chiral density ($OCD$) induced at any point in space by a given excitation \cite{chen2023observation,gryb2023two}:
\begin{equation}
{
 OCD=-\frac{\epsilon_0\omega}{2}\textit{Im}(\boldsymbol{E}^*\cdot \boldsymbol{B})
},
\label{equation2}
\end{equation}
where $\epsilon_0$ is the vacuum permittivity, $\omega$ is the angular frequency of light, \(\boldsymbol{E}^*\) is the complex conjugation of the electric field, and \(\boldsymbol{B}\) is the magnetic field. For plane waves, the \(\boldsymbol{E}\) and \(\boldsymbol{B}\) in free-space are orthogonal to each other, making the $OCD$ zero. However, the metasurface can support a nonzero $OCD$, provided that it has both in-plane and out-of-plane  broken symmetry \cite{zhang2022chiral,chen2023observation}. Slanted  nanostructures have been implemented to break both in-plane and out-of-plane symmetry to induce strong near-field chirality \cite{chen2023observation,zhang2022chiral}. Bilayers \cite{overvig2021chiral} and partially etched structures \cite{zhu2018giant} can also exhibit large optical chirality. However, the chiral responses of these metasurfaces are optimized for a certain angle of incidence, where the chirality at normal incidence is called intrinsic chirality and the chirality at oblique angle incidence is called extrinsic chirality. It is less challenging to obtain the extrinsic chirality as it does not necessarily require a broken symmetry along the out-of-plane direction. So far, none of them have shown the simultaneous chirality at the range of incident angles, i.e., normal and oblique incidence at the same time.

Exploiting the large number of degrees of freedom in a metasurface, various photonic band-structures have also been demonstrated recently \cite{nguyen2018symmetry}. Of particular interest, dispersion-free bands, so called ”flat” bands, have been raising particular interest. Due to its lack of dispersion, the flatband has extremely localized energy eigenstates and such flatband metasurfaces have been uti- lized as compact photodetectors \cite{choi2024nonlocal} and to enhance the emission from colloidal quantum wells \cite{munley2023visible}. In addition, this can also be utilized as narrowband color filter for large field of view augmented reality devices with minimized angular dispersion \cite{overvig2023zone}. To our knowledge, a flatband with chiral response has only been theoretically suggested, but and has not been demonstrated in photonic structures  \cite{sethi2024graph,ramachandran2017chiral,nakai2022perfect}.

In this study, we designed and experimentally demonstrated a metasurface with chiral flatband around the \(\Gamma\)-point with an incident angular range of $\sim 10^o$. We adopted an approach using partial etching and asymmetric meta-atoms to create the flatband with interaction between  propagating modes. The metasurface response can be described by a \(4\times4\) Hermitian Hamiltonian $H$ formed by the basis of  upper/lower and forward/backward propagating y-polarized modes along the x-direction \cite{nguyen2018symmetry}:
\begin{equation}
H = 
\begin{pmatrix}
w_1 + v_1k_x & U_1 & V_f & 0 \\
U_1 & w_1 - v_1k_x & 0 & V_f \\
V_f & 0 & w_2 + v_2k_x & \beta_2U_1 \\
0 & V_f & \beta_2U_1 & w_2 - v_2k_x
\end{pmatrix},
\label{equation3}
\end{equation}
where $U_1$ is the coupling rate between forward and backward propagating upper modes, $V_f$ is  the coupling rate between the upper and lower modes which propagates in the same direction, $\beta_2$ is the supplementary diffractive coupling factor. $w_{1,2}$ and $v_{1,2}$ are, respectively, the energies and the group velocities of the propagating waves under the $k_x$ vector coordinates. Since the patterned meta-atoms only exist in upper level due to the partial etching, the coupling rate between forward and backward propagating modes at the lower level as well as the supplementary diffraction coupling in the other direction are smaller and can be neglected in first approximation \cite{nguyen2018symmetry}. The asymmetry in coupling between orthogonal modes at the opposite ports can be optimized to achieve maximum far-field chiral response. We then optimize the band structure to achieve a flat band dispersion.

\begin{figure}[b]
\includegraphics{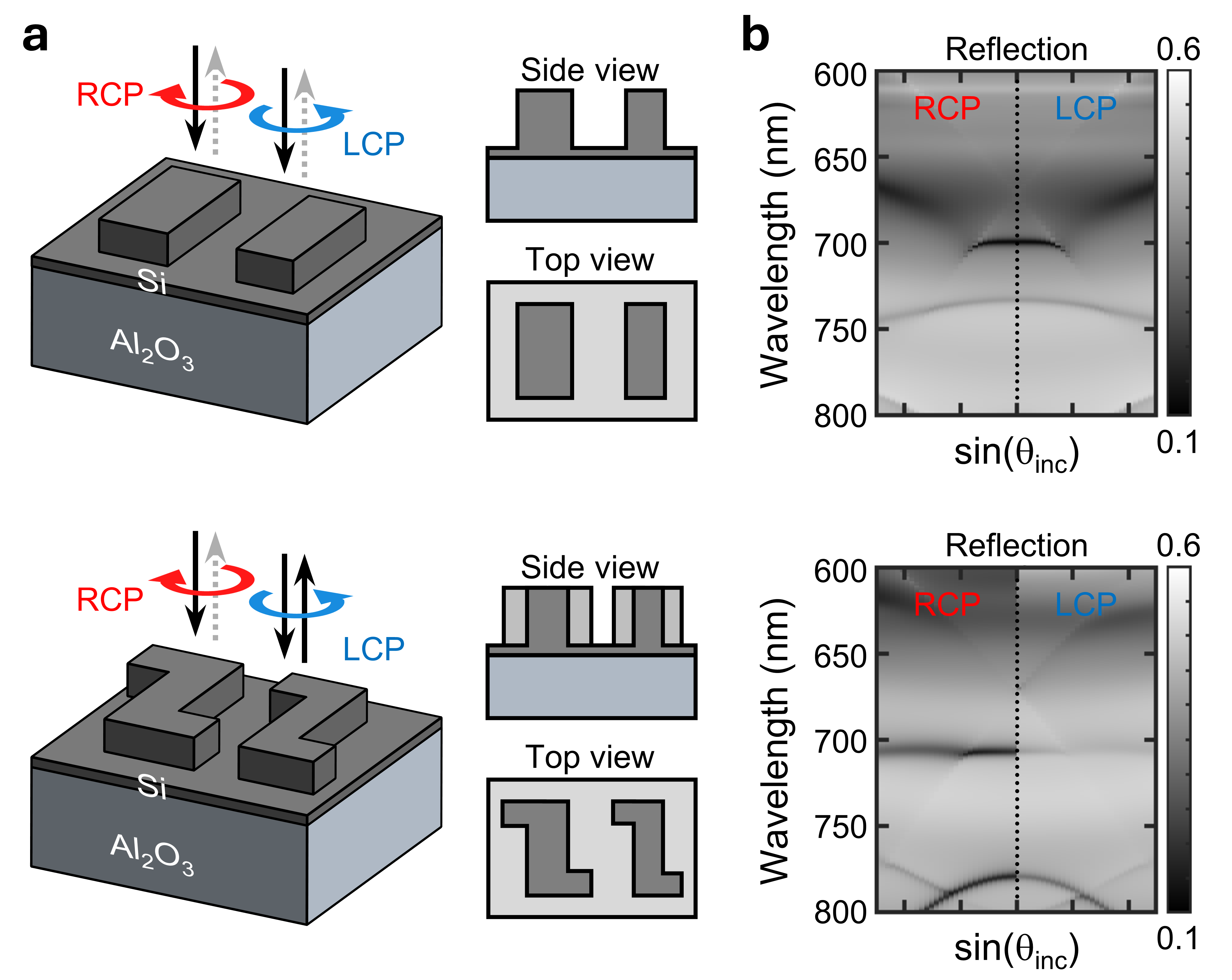}
\caption{\label{fig:Figure1} (a) Schematics of the 1D flatband metasurfaces with partial etching. (b) Simulated energy-momentum spectra of the metasurfaces in reflection configuration, where the RCP and LCP light is incident in oblique angle ($\theta_{inc}$). Incident angle varies from 0 to $15^o$.}
\end{figure}

To create a chiral metasurface, we first start with a one-dimensional (1D) high-contrast-grating structure, which supports a resonance for a particular linear polarization \cite{nguyen2018symmetry}. We then segment the 1D grating structure and break it down into two individual periodic meta-atoms (Fig. \ref{fig:Figure1}a) with size asymmetry. As there is a silicon slab underneath the arrays of meta-atoms due to partial etching, we can consider the structure as bilayer metasurface, whose response is optimized to support a flat- band response for a particular linear (TE) polarization. Fig. \ref{fig:Figure1}b shows the simulated reflection spectra of $RCP$ and $LCP$ incident light from the top which indicates no chirality, but a flat band dispersion around the $\Gamma$-point.

Given that the out-of-plane asymmetry is already introduced by the partial etching, we need an additional in-plane asymmetry to induce chirality. By implementing in-plane asymmetric "reverse-S shaped" chiral meta-atoms with partial etching, we achieve both in-plane and out-of-plane asymmetry, which are necessary to realize a nonzero $OCD$ (Fig. \ref{fig:Figure1}a). The chiral metasurface was optimized with finite-difference time- domain ($FDTD$) simulations, supporting a stark contrast in reflection between $RCP$ and $LCP$ inputs (Fig. \ref{fig:Figure1}b), while maintaining a flatband dispersion for an angular range of $\pm 5^o$.

\begin{figure}[b]
\includegraphics{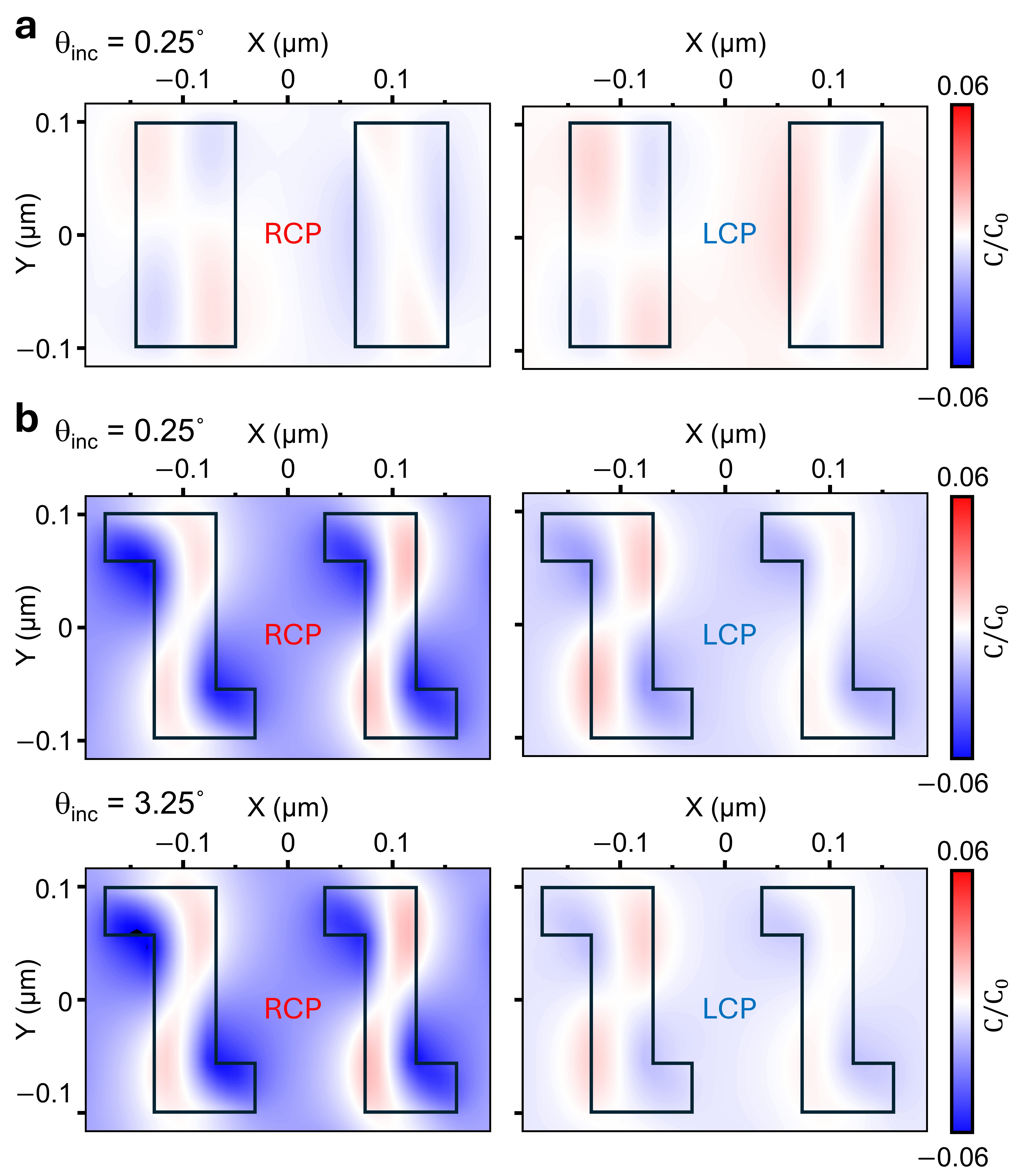}
\caption{\label{fig:Figure2} (a) Normalized near-field optical chiral density $OCD$ of the conventional flatband metasurface at resonance wavelength under RCP and LCP incidence light with $\theta_{inc} =0.25^o$). (b) Normalized near-field $OCD$ of the chiral flatband metasurface at the resonance wavelength under RCP and LCP incident light with $\theta_{inc} =0.25^o$ and $3.25^o$.}
\end{figure}

While the distinct contrast between $RCP$ and $LCP$ light reflection is calculated in the far-field, we can also relate this to the contrast in the near-field. The near-field $OCD$ is given by the imaginary parts of the product of $E^*$ and $B$. We normalized the OCD of the metasurface to the physical constant \(C_0=\epsilon_0\omega/2\), and plot its value at the resonance wavelength (Fig. \ref{fig:Figure2}). A conventional flatband metasurface consisting of 1D nano-bar does not support in-plane asymmetry, while out-of-plane asymmetry is realized by partial etching. Hence, we find no difference in near-field $OCD$ under $RCP$ and $LCP$ light illumination (Fig. \ref{fig:Figure2}a). Herein, we are looking at a slightly oblique incident angle ($\theta_{inc}=0.25^o$) which corresponds to the emergence of extrinsic chirality. On the other hand, as shown in Fig. \ref{fig:Figure2}b, when the meta-atoms have both in-plane and out-of-plane asymmetry, we find distinct $OCD$ under $RCP$ and $LCP$ light illumination. Similar to the far-field response, also the $OCD$ remains almost the same for different incident angles (up to an incident angle of $\pm5^o$), confirming the emergence of a chiral flatband.

In a chiral metasurface, both quality ($Q$) factor and $CD$ are important quantitative metrics. Various high-$Q$ chiral metasurfaces have been demonstrated, mostly leveraging quasi-bound states in the continuum (q-$BIC$) structures \cite{chen2023observation,zhang2022chiral,shi2022planar,gorkunov2020metasurfaces,overvig2021chiral}. Up to date, a chiral metasurface with slanted structures has reported the highest values of \(Q\sim 2,663\) and \(CD\sim 0.93\) (defined by its reflection contrast) \cite{chen2023observation}, but etching a material with a precisely oblique angle is not straightforward in conventional fabrication techniques. Additionally, the $Q$ and $CD$ were limited to a certain angle $\theta_{inc}$. In practice, due to the finite size of the structure and of the illumination, we should consider a range of $\theta_{inc}$, which leads to a reduction in $Q$ factor and $CD$. Our flatband chiral metasurface is also based on q-$BIC$ modes, which enable us to adjust both the $Q$ factor and $CD$ by modulating the structural parameters.
\begin{figure}[b]
\includegraphics{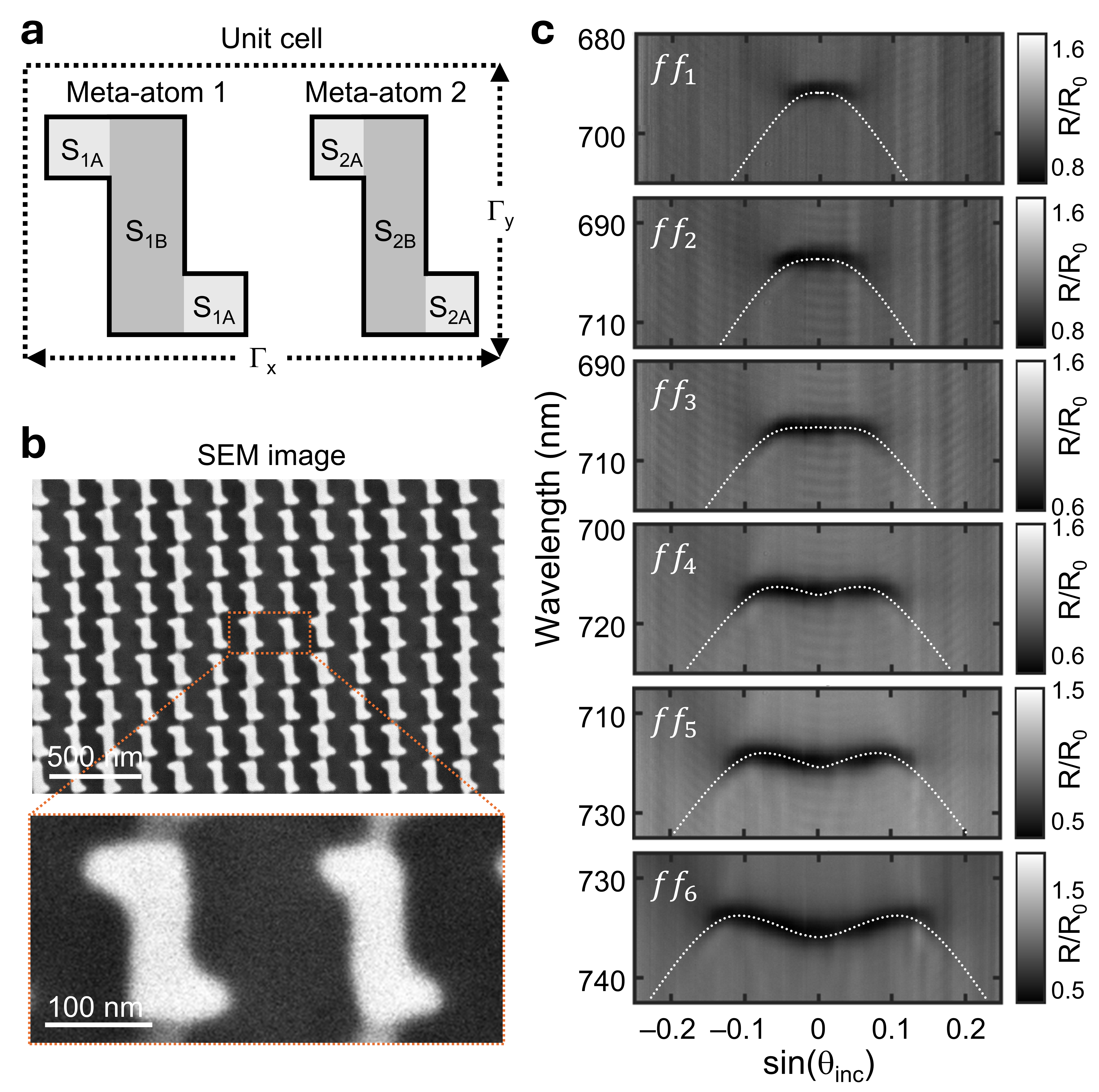}
\caption{\label{fig:Figure3} (a) Schematics of the meta-atoms within a unit cell of the chiral flatband metasurface. (b) Scanning electron microscopy images of the fabricated chiral metasurface. (c) Experimentally measured energy-momentum spectra of the chiral metasurface under $RCP$ light reflection with respect to the fill-factors $ff$. Dotted lines are the guide-to-eye of the photonic band structures evolving from parabolic to a flatband and multi-valley structures.}
\end{figure}
In order to control the $Q$ factor, we leverage the controlled asymmetry in the unit cell. Fig. \ref{fig:Figure3}a represents a unit cell of the chiral metasurface, consisting of two meta-atoms with an asymmetry factor ($\alpha$) defined as:
\begin{equation}
{
 \frac{1+\alpha}{1-\alpha} = \frac{S_1}{S_2} = \frac{(2S_{1A}+S_{1B})}{(2S_{2A}+S_{2B})}
},
\label{equation4}
\end{equation}
where $S_1$ and $S_2$ represent the area of the larger and smaller meta-atoms, respectively. Periods in $x$ and $y$ directions (\(\Gamma_x\) and \(\Gamma_y \)) define the unit cell size. Fill-factor ($ff$) defines the area of etched region within the unit cell:
\begin{equation}
{
 ff = 1-\frac{S_1 + S_2}{\Gamma_x \times \Gamma_y}
}.
\label{equation5}
\end{equation}
The partial etching ratio ($\epsilon$) is defined as the ratio between the etched depth and the total thickness (230 nm) of the crystalline silicon film. The chirality factor ($\eta$) defines the area ratio between the squares at two corners of each meta-atom and the total area:
\begin{equation}
{
 \eta = \frac{2S_{1A}}{S_1} = \frac{2S_{2A}}{S_2}
}.
\label{equation6}
\end{equation}
$\epsilon$ and $\eta$ contribute to the axial (out-of-plane) and lateral (in-plane) asymmetry, respectively, and we found an optimized set of parameters to maximize the contrast between the $RCP$ and $LCP$ reflection from the $FDTD$ simulation (details in Supporting Information). The asymmetry factor, $\alpha$, contributes to the $Q$ factor of the metasurface. As expected, the $Q$ increases as $\alpha$ decreases, and the response vanishes when $\alpha$ becomes zero, due to the $BIC$ feature \cite{choi2024nonlocal,wu2023giant}.

We fabricated the designed chiral metasurface in crystalline silicon on a sapphire substrate with a high-resolution electron-beam lithography using ZEP-520A resist, followed by development using amyl acetate, and reactive ion etching. The partial etching depth ($193\space nm$, which corresponds to $\epsilon=0.84$) was precisely controlled by the reactive ion etching time (see details in Supporting Information), while the etching rate is about $3\space nm/sec$. Fig. \ref{fig:Figure3}b shows scanning electron microscopy (SEM) images of the fabricated chiral metasurface, where we can clearly observe the "reverse-S shaped" meta-atoms.

We characterized the photonic band structure of the chiral metasurface using energy-momentum spectroscopy in reflection configuration (see details in Supporting Information). We put a Glan-Taylor polarizer and quarter-wave plate at the input light to create either $RCP$ or $LCP$ light, then measured the reflection in k-space and spectrally dispersed it with a blazed grating in spectrometer. Fig. \ref{fig:Figure3}c shows the energy-momentum spectra of the chiral metasurface with various fill-factors under $RCP$ light input, where the other structural parameters are the same. We can see the band structures evolving from parabolic to a flatband and multi-valley structures as the fill-factor increases \cite{nguyen2018symmetry, choi2024nonlocal}. Here, we subtracted the background noise and normalized with the reflection from the similar metasurface with zero asymmetry factor $(\alpha = 0)$, where the resonance disappeared as common BIC modes do, to normalize the spectral intensity profile of the input light and all the optics. From the energy-momentum spectra, we can clearly see the flatband which spreads over a wide range of angle ($\sim \pm 5^o$) at a certain fill-factor (Fig. \ref{fig:Figure3}c). All the energy-momentum spectra before normalization process is shown in the Supporting Information.

\begin{figure*}[t]
\includegraphics{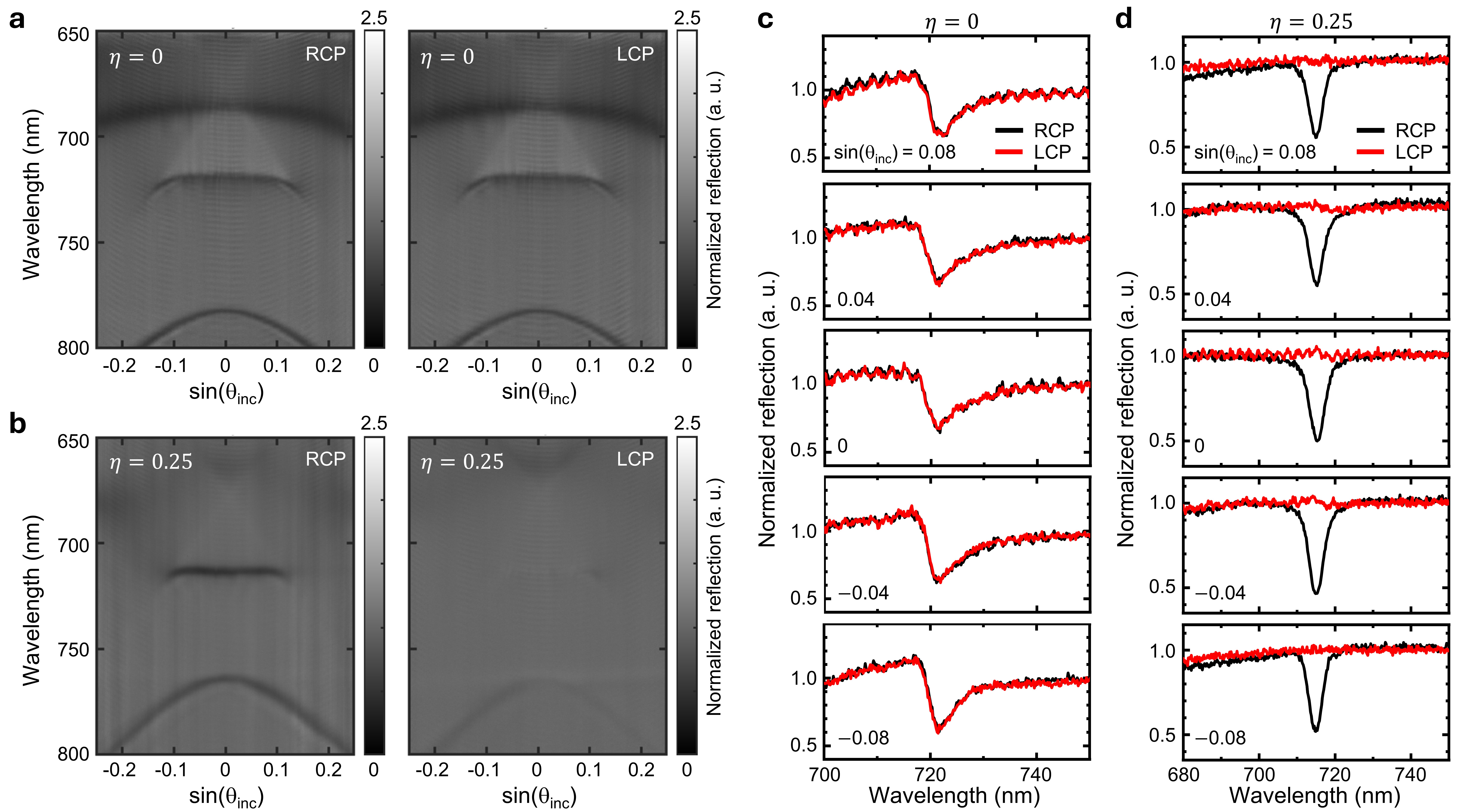}
\caption{\label{fig:Figure4} (a) Measured energy-momentum spectra of the metasurfaces with $\eta$ = 0 and $\alpha=0.10$ at $RCP$ and $LCP$ light input. (b) Measured energy-momentum spectra of the metasurfaces with $\eta$ = 0.25 and $\alpha=0.10$ at $RCP$ and $LCP$ light input. (c, d) Reflection spectra at different incident angles for $\eta$ = 0 and 0.25 metasurfaces at $RCP$ and $LCP$ light input. All the measured spectra were normalized by the results from the standard metasurface with zero asymmetry factor, while the other parameters are the same.}
\end{figure*}
 
To investigate the chirality of the metasurface with different structural chirality factors $\eta$ in far-field, we measured the energy-momentum spectra with $RCP$ and $LCP$ light input. Fig. \ref{fig:Figure4}a shows the energy-momentum spectra of the metasurfaces with $\eta$ = 0. We observe identical response for $RCP$ and $LCP$ light reflection, as expected from the near-field simulation results as shown in Fig. \ref{fig:Figure2}a. However, for the metasurface with $\eta$ = 0.25, we observed markedly different reflectivity under $RCP$ and $LCP$ illumination (Fig. \ref{fig:Figure4}a), consistent with the near-field simulation (Fig. \ref{fig:Figure2}b). As we described before, we normalized the reflection spectra with respect to the similar metasurface with zero asymmetry factor ($\alpha$ = 0)  while the other structural parameters are kept the same, to exclude all the spectral noise coming from the optical measurement systems (e.g, lens, mirrors, grating, and detector) and thin-film interference of the sample, and clearly focus on the photonic band structures. 

As our chiral metasurface also exhibits a flatband, we can observe a strong chirality for a relatively wide range of incident angles. Figs. \ref{fig:Figure4}c and d show the normalized reflection spectra at the various incident angles, $sin(\theta_{inc})=-0.08, -0.04, 0, 0.04$, and $0.08$, extracted from the energy-momentum spectra. We can see that the resonant wavelengths remain the same within $\sim \pm5^o$ of incident angles. The metasurface with zero chirality factor ($\eta = 0$) does not have any chirality, while the chiral metasurface with $\eta$ = 0.25 represents a strong chirality over a wide range of incident angles. 

\begin{figure*}[t]
\includegraphics{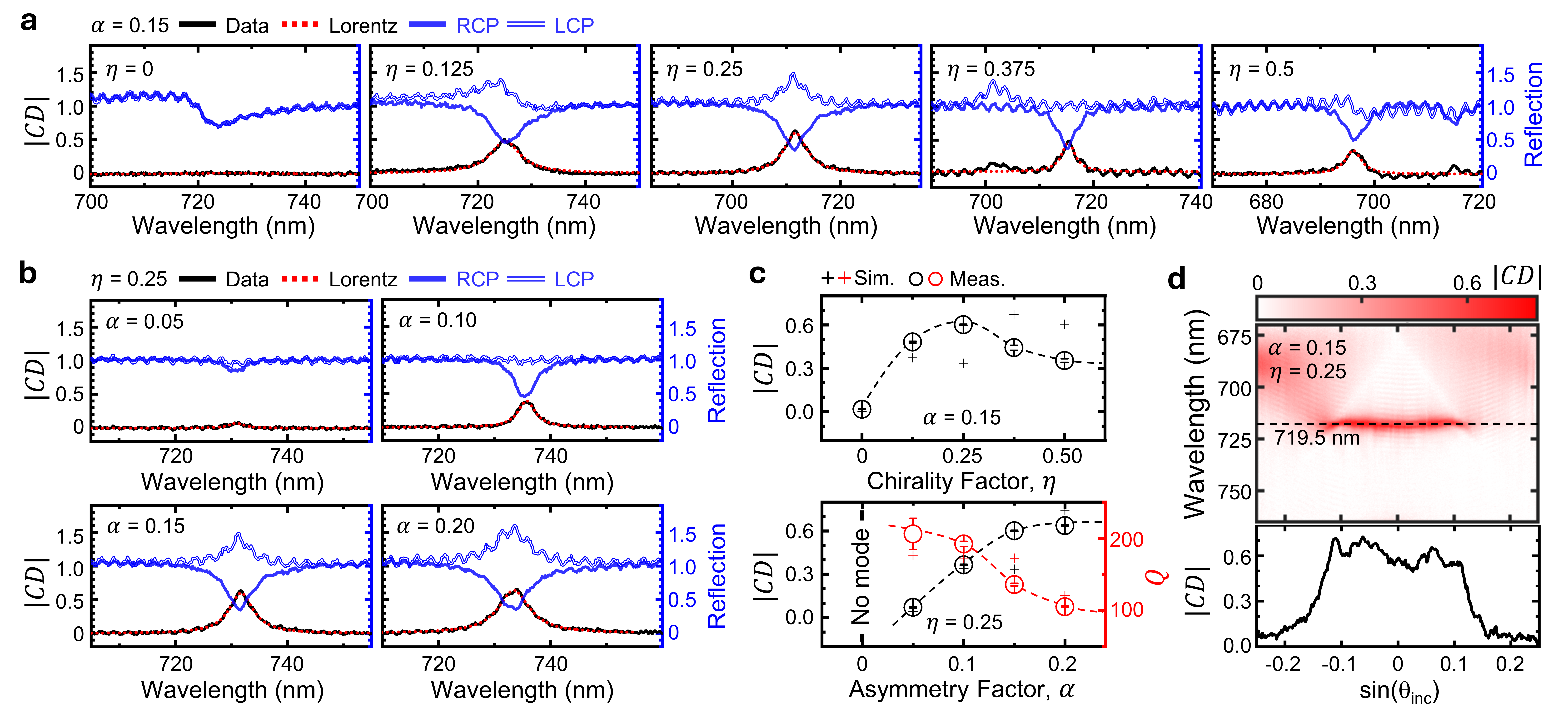}
\caption{\label{fig:Figure5} (a) Measured reflection spectra with $RCP$ and $LCP$ light input, and calculated $CD$ with Lorentzian fitting for different structural chirality factor, $\eta$, when $\alpha = 0.15$. (b) Measured reflection spectra with $RCP$ and $LCP$ light input, and calculated $CD$ with Lorentzian fitting for different asymmetry factor, $\alpha$, when $\eta = 0.25$. (c) Extracted maximum $CD$s (or $Q$ factors) at the resonance with respect to $\eta$ when $\alpha = 0.15$ and $\alpha$ when $\eta = 0.25$, respectively. Extracted $CD$ from the measurement (simulation) are shown in circles (crosses). Dashed line is the guide-to-eye of the $CD$s and $Q$ factors with respect to $\alpha$ and $\eta$. (d) Calculated angle-resolved $CD$ from the reflection measurement when $\alpha = 0.15$ and $\eta = 0.25$. Line-cut at the resonance wavelength ($\lambda=719.5\:nm$) shows a chiral flatband with large chirality ($CD>0.6$) over a wide range of $\theta_{inc}$ of $\sim\pm5^o$.}
\end{figure*}

We measured the energy-momentum spectra for several chiral flatband metasurfaces with different $\eta$ and $\alpha$, and observe good matching with the numerical simulation. Fig. \ref{fig:Figure5}a shows the measured reflection spectra with $RCP$ and $LCP$ light input, and the calculated $CD$ accordingly (as defined in Eq.~(\ref{equation1})) for the chiral metasurfaces with different structural chirality factors ($\eta$ = 0, 0.125, 0.25, 0.375, and 0.5) when $\alpha = 0.15$. From a Lorentzian fit, we can extract the $CD$ at the resonance which increases from zero and reaches maximum ($>0.60$) at around $\eta$ = 0.25, then decreases when $\eta$ is greater than 0.25. We can observe similar trends of $CD$ with respect to $\eta$ for different $\alpha$ values (details in Supporting Information). Fig. \ref{fig:Figure5}c shows the CD at the resonance with respect to $\eta$ when $\alpha = 0.15$ from both simulation (crosses) and experimental measurement (circles). Among a few representative $\eta$ values, we get the maximum value of $CD$ at $\eta=0.25$ in measurement, while at $\eta=0.375$ in simulation. We attribute this discrepancy to several factors: (1) partial etching depth uncertainty in reactive ion etching and (2) lateral size uncertainty in lithography. On the other hand, we can vary the asymmetry factor, $\alpha$, at certain chirality factor ($\eta=0.25$, in this case) and compare the reflection spectra at the resonance for $RCP$ and $LCP$ light input (Fig. \ref{fig:Figure5}b). We observe that the extracted far-field chiral dichroism $CD$ increases and $Q$ factor decreases at the same time when the $\alpha$ increases within our variable range, $\eta= 0 - 0.5$ and $\alpha = 0 - 0.2$ (Fig. \ref{fig:Figure5}c). This experimental finding (circles; black and red) also match the simulation (crosses; black and red). In addition, we showed a calculated angle-resolved CD from the reflection measurement under RCP and LCP light illumination in Fig. \ref{fig:Figure5}d, which ensures a relatively constant CD over a wide range of $\theta_{inc}$. We plot the calculated CD at the flatband resonance ($\lambda=719.5\:nm$), and show that a large chirality exists for a wide range of $\theta_{inc}$ of $\sim\pm5^o$. All the measured energy-momentum spectra with respect to various structural parameters are reported in the Supporting Information.

In conclusion, we designed and fabricated a flatband metasurface with large chirality ($CD > 0.6$) for a relatively wide range of incident angles ($\theta_{inc}\sim\pm5^o$). By controlling and optimizing the geometric parameters of the metasurface, we engineered the dispersion of light from parabolic to a flatband and multi-valley structures. In particular, a flatband metasurface can enhance nonlinear optical effects, thanks to the slow group velocity of light and large density of states. Imparting chiral response to such a metasurface opens up new opportunities to study light-matter interaction with emerging low-dimensional materials, such as transition metal dichalcogenide or ferromagnetic materials \cite{wang2022light,huang2017layer,fei2018two}. Specifically, we envision studying spin-selective valley polaritons and magneto-polaritons in such chiral flatband metasurface. Another important direction will be an angle-independent narrowband filter for augmented reality heads-up display system \cite{overvig2023zone,mao2016angle}. Both the narrow spectral window and large field of view can be achieved simultaneously with a dispersion-less flatband.

\begin{acknowledgments}
M.C. and A.M. were supported by the National Science Foundation Grant No. DMR-2019444 and NSF-210367. A.A. was supported by the Vannevar Bush Faculty Fellowship, the Air Force Office of Scientific Research MURI program and the Simons Foundation. Part of this work was conducted at the Washington Nanofabrication Facility/Molecular Analysis Facility, National Nanotechnology Coordinated Infrastructure (NNCI) site at the University of Washington with partial support from the National Science Foundation via awards NNCI-1542101 and NNCI-2025489.

\dots.
\end{acknowledgments}

\bibliography{apssamp}

\begin{thebibliography}{24}%
\makeatletter
\providecommand \@ifxundefined [1]{%
 \@ifx{#1\undefined}
}%
\providecommand \@ifnum [1]{%
 \ifnum #1\expandafter \@firstoftwo
 \else \expandafter \@secondoftwo
 \fi
}%
\providecommand \@ifx [1]{%
 \ifx #1\expandafter \@firstoftwo
 \else \expandafter \@secondoftwo
 \fi
}%
\providecommand \natexlab [1]{#1}%
\providecommand \enquote  [1]{``#1''}%
\providecommand \bibnamefont  [1]{#1}%
\providecommand \bibfnamefont [1]{#1}%
\providecommand \citenamefont [1]{#1}%
\providecommand \href@noop [0]{\@secondoftwo}%
\providecommand \href [0]{\begingroup \@sanitize@url \@href}%
\providecommand \@href[1]{\@@startlink{#1}\@@href}%
\providecommand \@@href[1]{\endgroup#1\@@endlink}%
\providecommand \@sanitize@url [0]{\catcode `\\12\catcode `\$12\catcode `\&12\catcode `\#12\catcode `\^12\catcode `\_12\catcode `\%12\relax}%
\providecommand \@@startlink[1]{}%
\providecommand \@@endlink[0]{}%
\providecommand \url  [0]{\begingroup\@sanitize@url \@url }%
\providecommand \@url [1]{\endgroup\@href {#1}{\urlprefix }}%
\providecommand \urlprefix  [0]{URL }%
\providecommand \Eprint [0]{\href }%
\providecommand \doibase [0]{https://doi.org/}%
\providecommand \selectlanguage [0]{\@gobble}%
\providecommand \bibinfo  [0]{\@secondoftwo}%
\providecommand \bibfield  [0]{\@secondoftwo}%
\providecommand \translation [1]{[#1]}%
\providecommand \BibitemOpen [0]{}%
\providecommand \bibitemStop [0]{}%
\providecommand \bibitemNoStop [0]{.\EOS\space}%
\providecommand \EOS [0]{\spacefactor3000\relax}%
\providecommand \BibitemShut  [1]{\csname bibitem#1\endcsname}%
\let\auto@bib@innerbib\@empty
\bibitem [{\citenamefont {Liu}\ \emph {et~al.}(2023)\citenamefont {Liu}, \citenamefont {Wu}, \citenamefont {Armstrong}, \citenamefont {Wolosker},\ and\ \citenamefont {Zheng}}]{liu2023detection}%
  \BibitemOpen
  \bibfield  {author} {\bibinfo {author} {\bibfnamefont {Y.}~\bibnamefont {Liu}}, \bibinfo {author} {\bibfnamefont {Z.}~\bibnamefont {Wu}}, \bibinfo {author} {\bibfnamefont {D.~W.}\ \bibnamefont {Armstrong}}, \bibinfo {author} {\bibfnamefont {H.}~\bibnamefont {Wolosker}},\ and\ \bibinfo {author} {\bibfnamefont {Y.}~\bibnamefont {Zheng}},\ }\bibfield  {title} {\bibinfo {title} {Detection and analysis of chiral molecules as disease biomarkers},\ }\href@noop {} {\bibfield  {journal} {\bibinfo  {journal} {Nature Reviews Chemistry}\ }\textbf {\bibinfo {volume} {7}},\ \bibinfo {pages} {355} (\bibinfo {year} {2023})}\BibitemShut {NoStop}%
\bibitem [{\citenamefont {Schilthuizen}\ and\ \citenamefont {Davison}(2005)}]{schilthuizen2005convoluted}%
  \BibitemOpen
  \bibfield  {author} {\bibinfo {author} {\bibfnamefont {M.}~\bibnamefont {Schilthuizen}}\ and\ \bibinfo {author} {\bibfnamefont {A.}~\bibnamefont {Davison}},\ }\bibfield  {title} {\bibinfo {title} {The convoluted evolution of snail chirality},\ }\href@noop {} {\bibfield  {journal} {\bibinfo  {journal} {Naturwissenschaften}\ }\textbf {\bibinfo {volume} {92}},\ \bibinfo {pages} {504} (\bibinfo {year} {2005})}\BibitemShut {NoStop}%
\bibitem [{\citenamefont {Cook}(1903)}]{cook1903spirals}%
  \BibitemOpen
  \bibfield  {author} {\bibinfo {author} {\bibfnamefont {T.~A.}\ \bibnamefont {Cook}},\ }\bibfield  {title} {\bibinfo {title} {Spirals in nature and art},\ }\href@noop {} {\bibfield  {journal} {\bibinfo  {journal} {Nature}\ }\textbf {\bibinfo {volume} {68}},\ \bibinfo {pages} {296} (\bibinfo {year} {1903})}\BibitemShut {NoStop}%
\bibitem [{\citenamefont {Tang}\ and\ \citenamefont {Cohen}(2010)}]{tang2010optical}%
  \BibitemOpen
  \bibfield  {author} {\bibinfo {author} {\bibfnamefont {Y.}~\bibnamefont {Tang}}\ and\ \bibinfo {author} {\bibfnamefont {A.~E.}\ \bibnamefont {Cohen}},\ }\bibfield  {title} {\bibinfo {title} {Optical chirality and its interaction with matter},\ }\href@noop {} {\bibfield  {journal} {\bibinfo  {journal} {Physical review letters}\ }\textbf {\bibinfo {volume} {104}},\ \bibinfo {pages} {163901} (\bibinfo {year} {2010})}\BibitemShut {NoStop}%
\bibitem [{\citenamefont {Chen}\ \emph {et~al.}(2023)\citenamefont {Chen}, \citenamefont {Deng}, \citenamefont {Sha}, \citenamefont {Chen}, \citenamefont {Wang}, \citenamefont {Chen}, \citenamefont {Wu}, \citenamefont {Chu}, \citenamefont {Kivshar}, \citenamefont {Xiao} \emph {et~al.}}]{chen2023observation}%
  \BibitemOpen
  \bibfield  {author} {\bibinfo {author} {\bibfnamefont {Y.}~\bibnamefont {Chen}}, \bibinfo {author} {\bibfnamefont {H.}~\bibnamefont {Deng}}, \bibinfo {author} {\bibfnamefont {X.}~\bibnamefont {Sha}}, \bibinfo {author} {\bibfnamefont {W.}~\bibnamefont {Chen}}, \bibinfo {author} {\bibfnamefont {R.}~\bibnamefont {Wang}}, \bibinfo {author} {\bibfnamefont {Y.-H.}\ \bibnamefont {Chen}}, \bibinfo {author} {\bibfnamefont {D.}~\bibnamefont {Wu}}, \bibinfo {author} {\bibfnamefont {J.}~\bibnamefont {Chu}}, \bibinfo {author} {\bibfnamefont {Y.~S.}\ \bibnamefont {Kivshar}}, \bibinfo {author} {\bibfnamefont {S.}~\bibnamefont {Xiao}}, \emph {et~al.},\ }\bibfield  {title} {\bibinfo {title} {Observation of intrinsic chiral bound states in the continuum},\ }\href@noop {} {\bibfield  {journal} {\bibinfo  {journal} {Nature}\ }\textbf {\bibinfo {volume} {613}},\ \bibinfo {pages} {474} (\bibinfo {year} {2023})}\BibitemShut {NoStop}%
\bibitem [{\citenamefont {Glover}\ \emph {et~al.}(1985)\citenamefont {Glover}, \citenamefont {Hayes}, \citenamefont {Pelc}, \citenamefont {Edelstein}, \citenamefont {Mueller}, \citenamefont {Hart}, \citenamefont {Hardy}, \citenamefont {O'donnell},\ and\ \citenamefont {Barber}}]{glover1985comparison}%
  \BibitemOpen
  \bibfield  {author} {\bibinfo {author} {\bibfnamefont {G.}~\bibnamefont {Glover}}, \bibinfo {author} {\bibfnamefont {C.}~\bibnamefont {Hayes}}, \bibinfo {author} {\bibfnamefont {N.}~\bibnamefont {Pelc}}, \bibinfo {author} {\bibfnamefont {W.}~\bibnamefont {Edelstein}}, \bibinfo {author} {\bibfnamefont {O.}~\bibnamefont {Mueller}}, \bibinfo {author} {\bibfnamefont {H.}~\bibnamefont {Hart}}, \bibinfo {author} {\bibfnamefont {C.}~\bibnamefont {Hardy}}, \bibinfo {author} {\bibfnamefont {M.}~\bibnamefont {O'donnell}},\ and\ \bibinfo {author} {\bibfnamefont {W.}~\bibnamefont {Barber}},\ }\bibfield  {title} {\bibinfo {title} {Comparison of linear and circular polarization for magnetic resonance imaging},\ }\href@noop {} {\bibfield  {journal} {\bibinfo  {journal} {Journal of Magnetic Resonance (1969)}\ }\textbf {\bibinfo {volume} {64}},\ \bibinfo {pages} {255} (\bibinfo {year} {1985})}\BibitemShut {NoStop}%
\bibitem [{\citenamefont {Zhang}\ \emph {et~al.}(2022)\citenamefont {Zhang}, \citenamefont {Liu}, \citenamefont {Han}, \citenamefont {Kivshar},\ and\ \citenamefont {Song}}]{zhang2022chiral}%
  \BibitemOpen
  \bibfield  {author} {\bibinfo {author} {\bibfnamefont {X.}~\bibnamefont {Zhang}}, \bibinfo {author} {\bibfnamefont {Y.}~\bibnamefont {Liu}}, \bibinfo {author} {\bibfnamefont {J.}~\bibnamefont {Han}}, \bibinfo {author} {\bibfnamefont {Y.}~\bibnamefont {Kivshar}},\ and\ \bibinfo {author} {\bibfnamefont {Q.}~\bibnamefont {Song}},\ }\bibfield  {title} {\bibinfo {title} {Chiral emission from resonant metasurfaces},\ }\href@noop {} {\bibfield  {journal} {\bibinfo  {journal} {Science}\ }\textbf {\bibinfo {volume} {377}},\ \bibinfo {pages} {1215} (\bibinfo {year} {2022})}\BibitemShut {NoStop}%
\bibitem [{\citenamefont {Gryb}\ \emph {et~al.}(2023)\citenamefont {Gryb}, \citenamefont {Wendisch}, \citenamefont {Aigner}, \citenamefont {G{\``o}lz}, \citenamefont {Tittl}, \citenamefont {de~S.~Menezes},\ and\ \citenamefont {Maier}}]{gryb2023two}%
  \BibitemOpen
  \bibfield  {author} {\bibinfo {author} {\bibfnamefont {D.}~\bibnamefont {Gryb}}, \bibinfo {author} {\bibfnamefont {F.~J.}\ \bibnamefont {Wendisch}}, \bibinfo {author} {\bibfnamefont {A.}~\bibnamefont {Aigner}}, \bibinfo {author} {\bibfnamefont {T.}~\bibnamefont {G{\``o}lz}}, \bibinfo {author} {\bibfnamefont {A.}~\bibnamefont {Tittl}}, \bibinfo {author} {\bibfnamefont {L.}~\bibnamefont {de~S.~Menezes}},\ and\ \bibinfo {author} {\bibfnamefont {S.~A.}\ \bibnamefont {Maier}},\ }\bibfield  {title} {\bibinfo {title} {Two-dimensional chiral metasurfaces obtained by geometrically simple meta-atom rotations},\ }\href@noop {} {\bibfield  {journal} {\bibinfo  {journal} {Nano Letters}\ }\textbf {\bibinfo {volume} {23}},\ \bibinfo {pages} {8891} (\bibinfo {year} {2023})}\BibitemShut {NoStop}%
\bibitem [{\citenamefont {Overvig}\ \emph {et~al.}(2021)\citenamefont {Overvig}, \citenamefont {Yu},\ and\ \citenamefont {Al{\`u}}}]{overvig2021chiral}%
  \BibitemOpen
  \bibfield  {author} {\bibinfo {author} {\bibfnamefont {A.}~\bibnamefont {Overvig}}, \bibinfo {author} {\bibfnamefont {N.}~\bibnamefont {Yu}},\ and\ \bibinfo {author} {\bibfnamefont {A.}~\bibnamefont {Al{\`u}}},\ }\bibfield  {title} {\bibinfo {title} {Chiral quasi-bound states in the continuum},\ }\href@noop {} {\bibfield  {journal} {\bibinfo  {journal} {Physical Review Letters}\ }\textbf {\bibinfo {volume} {126}},\ \bibinfo {pages} {073001} (\bibinfo {year} {2021})}\BibitemShut {NoStop}%
\bibitem [{\citenamefont {Zhu}\ \emph {et~al.}(2018)\citenamefont {Zhu}, \citenamefont {Chen}, \citenamefont {Zaidi}, \citenamefont {Huang}, \citenamefont {Khorasaninejad}, \citenamefont {Sanjeev}, \citenamefont {Qiu},\ and\ \citenamefont {Capasso}}]{zhu2018giant}%
  \BibitemOpen
  \bibfield  {author} {\bibinfo {author} {\bibfnamefont {A.~Y.}\ \bibnamefont {Zhu}}, \bibinfo {author} {\bibfnamefont {W.~T.}\ \bibnamefont {Chen}}, \bibinfo {author} {\bibfnamefont {A.}~\bibnamefont {Zaidi}}, \bibinfo {author} {\bibfnamefont {Y.-W.}\ \bibnamefont {Huang}}, \bibinfo {author} {\bibfnamefont {M.}~\bibnamefont {Khorasaninejad}}, \bibinfo {author} {\bibfnamefont {V.}~\bibnamefont {Sanjeev}}, \bibinfo {author} {\bibfnamefont {C.-W.}\ \bibnamefont {Qiu}},\ and\ \bibinfo {author} {\bibfnamefont {F.}~\bibnamefont {Capasso}},\ }\bibfield  {title} {\bibinfo {title} {Giant intrinsic chiro-optical activity in planar dielectric nanostructures},\ }\href@noop {} {\bibfield  {journal} {\bibinfo  {journal} {Light: Science \& Applications}\ }\textbf {\bibinfo {volume} {7}},\ \bibinfo {pages} {17158} (\bibinfo {year} {2018})}\BibitemShut {NoStop}%
\bibitem [{\citenamefont {Nguyen}\ \emph {et~al.}(2018)\citenamefont {Nguyen}, \citenamefont {Dubois}, \citenamefont {Deschamps}, \citenamefont {Cueff}, \citenamefont {Pardon}, \citenamefont {Leclercq}, \citenamefont {Seassal}, \citenamefont {Letartre},\ and\ \citenamefont {Viktorovitch}}]{nguyen2018symmetry}%
  \BibitemOpen
  \bibfield  {author} {\bibinfo {author} {\bibfnamefont {H.~S.}\ \bibnamefont {Nguyen}}, \bibinfo {author} {\bibfnamefont {F.}~\bibnamefont {Dubois}}, \bibinfo {author} {\bibfnamefont {T.}~\bibnamefont {Deschamps}}, \bibinfo {author} {\bibfnamefont {S.}~\bibnamefont {Cueff}}, \bibinfo {author} {\bibfnamefont {A.}~\bibnamefont {Pardon}}, \bibinfo {author} {\bibfnamefont {J.-L.}\ \bibnamefont {Leclercq}}, \bibinfo {author} {\bibfnamefont {C.}~\bibnamefont {Seassal}}, \bibinfo {author} {\bibfnamefont {X.}~\bibnamefont {Letartre}},\ and\ \bibinfo {author} {\bibfnamefont {P.}~\bibnamefont {Viktorovitch}},\ }\bibfield  {title} {\bibinfo {title} {Symmetry breaking in photonic crystals: On-demand dispersion from flatband to dirac cones},\ }\href@noop {} {\bibfield  {journal} {\bibinfo  {journal} {Physical review letters}\ }\textbf {\bibinfo {volume} {120}},\ \bibinfo {pages} {066102} (\bibinfo {year} {2018})}\BibitemShut {NoStop}%
\bibitem [{\citenamefont {Choi}\ \emph {et~al.}(2024)\citenamefont {Choi}, \citenamefont {Munley}, \citenamefont {Froch}, \citenamefont {Chen},\ and\ \citenamefont {Majumdar}}]{choi2024nonlocal}%
  \BibitemOpen
  \bibfield  {author} {\bibinfo {author} {\bibfnamefont {M.}~\bibnamefont {Choi}}, \bibinfo {author} {\bibfnamefont {C.}~\bibnamefont {Munley}}, \bibinfo {author} {\bibfnamefont {J.~E.}\ \bibnamefont {Froch}}, \bibinfo {author} {\bibfnamefont {R.}~\bibnamefont {Chen}},\ and\ \bibinfo {author} {\bibfnamefont {A.}~\bibnamefont {Majumdar}},\ }\bibfield  {title} {\bibinfo {title} {Nonlocal, flat-band meta-optics for monolithic, high-efficiency, compact photodetectors},\ }\href@noop {} {\bibfield  {journal} {\bibinfo  {journal} {Nano Letters}\ }\textbf {\bibinfo {volume} {24}},\ \bibinfo {pages} {3150} (\bibinfo {year} {2024})}\BibitemShut {NoStop}%
\bibitem [{\citenamefont {Munley}\ \emph {et~al.}(2023)\citenamefont {Munley}, \citenamefont {Manna}, \citenamefont {Sharp}, \citenamefont {Choi}, \citenamefont {Nguyen}, \citenamefont {Cossairt}, \citenamefont {Li}, \citenamefont {Barnard},\ and\ \citenamefont {Majumdar}}]{munley2023visible}%
  \BibitemOpen
  \bibfield  {author} {\bibinfo {author} {\bibfnamefont {C.}~\bibnamefont {Munley}}, \bibinfo {author} {\bibfnamefont {A.}~\bibnamefont {Manna}}, \bibinfo {author} {\bibfnamefont {D.}~\bibnamefont {Sharp}}, \bibinfo {author} {\bibfnamefont {M.}~\bibnamefont {Choi}}, \bibinfo {author} {\bibfnamefont {H.~A.}\ \bibnamefont {Nguyen}}, \bibinfo {author} {\bibfnamefont {B.~M.}\ \bibnamefont {Cossairt}}, \bibinfo {author} {\bibfnamefont {M.}~\bibnamefont {Li}}, \bibinfo {author} {\bibfnamefont {A.~W.}\ \bibnamefont {Barnard}},\ and\ \bibinfo {author} {\bibfnamefont {A.}~\bibnamefont {Majumdar}},\ }\bibfield  {title} {\bibinfo {title} {Visible wavelength flatband in a gallium phosphide metasurface},\ }\href@noop {} {\bibfield  {journal} {\bibinfo  {journal} {ACS Photonics}\ }\textbf {\bibinfo {volume} {10}},\ \bibinfo {pages} {2456} (\bibinfo {year} {2023})}\BibitemShut {NoStop}%
\bibitem [{\citenamefont {Overvig}\ \emph {et~al.}(2023)\citenamefont {Overvig}, \citenamefont {Cotrufo}, \citenamefont {Markowitz}, \citenamefont {Zhou}, \citenamefont {Hao}, \citenamefont {Stensvad}, \citenamefont {Schardt},\ and\ \citenamefont {Al{\`u}}}]{overvig2023zone}%
  \BibitemOpen
  \bibfield  {author} {\bibinfo {author} {\bibfnamefont {A.~C.}\ \bibnamefont {Overvig}}, \bibinfo {author} {\bibfnamefont {M.}~\bibnamefont {Cotrufo}}, \bibinfo {author} {\bibfnamefont {M.}~\bibnamefont {Markowitz}}, \bibinfo {author} {\bibfnamefont {Y.}~\bibnamefont {Zhou}}, \bibinfo {author} {\bibfnamefont {B.}~\bibnamefont {Hao}}, \bibinfo {author} {\bibfnamefont {K.}~\bibnamefont {Stensvad}}, \bibinfo {author} {\bibfnamefont {C.}~\bibnamefont {Schardt}},\ and\ \bibinfo {author} {\bibfnamefont {A.}~\bibnamefont {Al{\`u}}},\ }\bibfield  {title} {\bibinfo {title} {Zone-folded quasi-bound state metasurfaces with customized, symmetry-protected energy-momentum relations},\ }\href@noop {} {\bibfield  {journal} {\bibinfo  {journal} {ACS Photonics}\ }\textbf {\bibinfo {volume} {10}},\ \bibinfo {pages} {1832} (\bibinfo {year} {2023})}\BibitemShut {NoStop}%
\bibitem [{\citenamefont {Sethi}\ \emph {et~al.}(2024)\citenamefont {Sethi}, \citenamefont {Xia}, \citenamefont {Kim}, \citenamefont {Liu}, \citenamefont {Li},\ and\ \citenamefont {Liu}}]{sethi2024graph}%
  \BibitemOpen
  \bibfield  {author} {\bibinfo {author} {\bibfnamefont {G.}~\bibnamefont {Sethi}}, \bibinfo {author} {\bibfnamefont {B.}~\bibnamefont {Xia}}, \bibinfo {author} {\bibfnamefont {D.}~\bibnamefont {Kim}}, \bibinfo {author} {\bibfnamefont {H.}~\bibnamefont {Liu}}, \bibinfo {author} {\bibfnamefont {X.}~\bibnamefont {Li}},\ and\ \bibinfo {author} {\bibfnamefont {F.}~\bibnamefont {Liu}},\ }\bibfield  {title} {\bibinfo {title} {Graph theorem for chiral exact flat bands at charge neutrality},\ }\href@noop {} {\bibfield  {journal} {\bibinfo  {journal} {Physical Review B}\ }\textbf {\bibinfo {volume} {109}},\ \bibinfo {pages} {035140} (\bibinfo {year} {2024})}\BibitemShut {NoStop}%
\bibitem [{\citenamefont {Ramachandran}\ \emph {et~al.}(2017)\citenamefont {Ramachandran}, \citenamefont {Andreanov},\ and\ \citenamefont {Flach}}]{ramachandran2017chiral}%
  \BibitemOpen
  \bibfield  {author} {\bibinfo {author} {\bibfnamefont {A.}~\bibnamefont {Ramachandran}}, \bibinfo {author} {\bibfnamefont {A.}~\bibnamefont {Andreanov}},\ and\ \bibinfo {author} {\bibfnamefont {S.}~\bibnamefont {Flach}},\ }\bibfield  {title} {\bibinfo {title} {Chiral flat bands: Existence, engineering, and stability},\ }\href@noop {} {\bibfield  {journal} {\bibinfo  {journal} {Physical Review B}\ }\textbf {\bibinfo {volume} {96}},\ \bibinfo {pages} {161104} (\bibinfo {year} {2017})}\BibitemShut {NoStop}%
\bibitem [{\citenamefont {Nakai}\ and\ \citenamefont {Hotta}(2022)}]{nakai2022perfect}%
  \BibitemOpen
  \bibfield  {author} {\bibinfo {author} {\bibfnamefont {H.}~\bibnamefont {Nakai}}\ and\ \bibinfo {author} {\bibfnamefont {C.}~\bibnamefont {Hotta}},\ }\bibfield  {title} {\bibinfo {title} {Perfect flat band with chirality and charge ordering out of strong spin-orbit interaction},\ }\href@noop {} {\bibfield  {journal} {\bibinfo  {journal} {Nature Communications}\ }\textbf {\bibinfo {volume} {13}},\ \bibinfo {pages} {579} (\bibinfo {year} {2022})}\BibitemShut {NoStop}%
\bibitem [{\citenamefont {Shi}\ \emph {et~al.}(2022)\citenamefont {Shi}, \citenamefont {Deng}, \citenamefont {Geng}, \citenamefont {Zeng}, \citenamefont {Zeng}, \citenamefont {Hu}, \citenamefont {Overvig}, \citenamefont {Li}, \citenamefont {Qiu}, \citenamefont {Al{\`u}} \emph {et~al.}}]{shi2022planar}%
  \BibitemOpen
  \bibfield  {author} {\bibinfo {author} {\bibfnamefont {T.}~\bibnamefont {Shi}}, \bibinfo {author} {\bibfnamefont {Z.-L.}\ \bibnamefont {Deng}}, \bibinfo {author} {\bibfnamefont {G.}~\bibnamefont {Geng}}, \bibinfo {author} {\bibfnamefont {X.}~\bibnamefont {Zeng}}, \bibinfo {author} {\bibfnamefont {Y.}~\bibnamefont {Zeng}}, \bibinfo {author} {\bibfnamefont {G.}~\bibnamefont {Hu}}, \bibinfo {author} {\bibfnamefont {A.}~\bibnamefont {Overvig}}, \bibinfo {author} {\bibfnamefont {J.}~\bibnamefont {Li}}, \bibinfo {author} {\bibfnamefont {C.-W.}\ \bibnamefont {Qiu}}, \bibinfo {author} {\bibfnamefont {A.}~\bibnamefont {Al{\`u}}}, \emph {et~al.},\ }\bibfield  {title} {\bibinfo {title} {Planar chiral metasurfaces with maximal and tunable chiroptical response driven by bound states in the continuum},\ }\href@noop {} {\bibfield  {journal} {\bibinfo  {journal} {Nature Communications}\ }\textbf {\bibinfo {volume} {13}},\ \bibinfo {pages} {4111} (\bibinfo {year} {2022})}\BibitemShut {NoStop}%
\bibitem [{\citenamefont {Gorkunov}\ \emph {et~al.}(2020)\citenamefont {Gorkunov}, \citenamefont {Antonov},\ and\ \citenamefont {Kivshar}}]{gorkunov2020metasurfaces}%
  \BibitemOpen
  \bibfield  {author} {\bibinfo {author} {\bibfnamefont {M.~V.}\ \bibnamefont {Gorkunov}}, \bibinfo {author} {\bibfnamefont {A.~A.}\ \bibnamefont {Antonov}},\ and\ \bibinfo {author} {\bibfnamefont {Y.~S.}\ \bibnamefont {Kivshar}},\ }\bibfield  {title} {\bibinfo {title} {Metasurfaces with maximum chirality empowered by bound states in the continuum},\ }\href@noop {} {\bibfield  {journal} {\bibinfo  {journal} {Physical Review Letters}\ }\textbf {\bibinfo {volume} {125}},\ \bibinfo {pages} {093903} (\bibinfo {year} {2020})}\BibitemShut {NoStop}%
\bibitem [{\citenamefont {Wu}\ \emph {et~al.}(2023)\citenamefont {Wu}, \citenamefont {Liu}, \citenamefont {Long}, \citenamefont {Xiao},\ and\ \citenamefont {Chen}}]{wu2023giant}%
  \BibitemOpen
  \bibfield  {author} {\bibinfo {author} {\bibfnamefont {F.}~\bibnamefont {Wu}}, \bibinfo {author} {\bibfnamefont {T.}~\bibnamefont {Liu}}, \bibinfo {author} {\bibfnamefont {Y.}~\bibnamefont {Long}}, \bibinfo {author} {\bibfnamefont {S.}~\bibnamefont {Xiao}},\ and\ \bibinfo {author} {\bibfnamefont {G.}~\bibnamefont {Chen}},\ }\bibfield  {title} {\bibinfo {title} {Giant photonic spin hall effect empowered by polarization-dependent quasibound states in the continuum in compound grating waveguide structures},\ }\href@noop {} {\bibfield  {journal} {\bibinfo  {journal} {Physical Review B}\ }\textbf {\bibinfo {volume} {107}},\ \bibinfo {pages} {165428} (\bibinfo {year} {2023})}\BibitemShut {NoStop}%
\bibitem [{\citenamefont {Wang}\ \emph {et~al.}(2022)\citenamefont {Wang}, \citenamefont {Xiao}, \citenamefont {Park}, \citenamefont {Zhu}, \citenamefont {Wang}, \citenamefont {Taniguchi}, \citenamefont {Watanabe}, \citenamefont {Yan}, \citenamefont {Xiao}, \citenamefont {Gamelin} \emph {et~al.}}]{wang2022light}%
  \BibitemOpen
  \bibfield  {author} {\bibinfo {author} {\bibfnamefont {X.}~\bibnamefont {Wang}}, \bibinfo {author} {\bibfnamefont {C.}~\bibnamefont {Xiao}}, \bibinfo {author} {\bibfnamefont {H.}~\bibnamefont {Park}}, \bibinfo {author} {\bibfnamefont {J.}~\bibnamefont {Zhu}}, \bibinfo {author} {\bibfnamefont {C.}~\bibnamefont {Wang}}, \bibinfo {author} {\bibfnamefont {T.}~\bibnamefont {Taniguchi}}, \bibinfo {author} {\bibfnamefont {K.}~\bibnamefont {Watanabe}}, \bibinfo {author} {\bibfnamefont {J.}~\bibnamefont {Yan}}, \bibinfo {author} {\bibfnamefont {D.}~\bibnamefont {Xiao}}, \bibinfo {author} {\bibfnamefont {D.~R.}\ \bibnamefont {Gamelin}}, \emph {et~al.},\ }\bibfield  {title} {\bibinfo {title} {Light-induced ferromagnetism in moir{\'e} superlattices},\ }\href@noop {} {\bibfield  {journal} {\bibinfo  {journal} {Nature}\ }\textbf {\bibinfo {volume} {604}},\ \bibinfo {pages} {468} (\bibinfo {year} {2022})}\BibitemShut {NoStop}%
\bibitem [{\citenamefont {Huang}\ \emph {et~al.}(2017)\citenamefont {Huang}, \citenamefont {Clark}, \citenamefont {Navarro-Moratalla}, \citenamefont {Klein}, \citenamefont {Cheng}, \citenamefont {Seyler}, \citenamefont {Zhong}, \citenamefont {Schmidgall}, \citenamefont {McGuire}, \citenamefont {Cobden} \emph {et~al.}}]{huang2017layer}%
  \BibitemOpen
  \bibfield  {author} {\bibinfo {author} {\bibfnamefont {B.}~\bibnamefont {Huang}}, \bibinfo {author} {\bibfnamefont {G.}~\bibnamefont {Clark}}, \bibinfo {author} {\bibfnamefont {E.}~\bibnamefont {Navarro-Moratalla}}, \bibinfo {author} {\bibfnamefont {D.~R.}\ \bibnamefont {Klein}}, \bibinfo {author} {\bibfnamefont {R.}~\bibnamefont {Cheng}}, \bibinfo {author} {\bibfnamefont {K.~L.}\ \bibnamefont {Seyler}}, \bibinfo {author} {\bibfnamefont {D.}~\bibnamefont {Zhong}}, \bibinfo {author} {\bibfnamefont {E.}~\bibnamefont {Schmidgall}}, \bibinfo {author} {\bibfnamefont {M.~A.}\ \bibnamefont {McGuire}}, \bibinfo {author} {\bibfnamefont {D.~H.}\ \bibnamefont {Cobden}}, \emph {et~al.},\ }\bibfield  {title} {\bibinfo {title} {Layer-dependent ferromagnetism in a van der waals crystal down to the monolayer limit},\ }\href@noop {} {\bibfield  {journal} {\bibinfo  {journal} {Nature}\ }\textbf {\bibinfo {volume} {546}},\ \bibinfo {pages} {270} (\bibinfo {year} {2017})}\BibitemShut {NoStop}%
\bibitem [{\citenamefont {Fei}\ \emph {et~al.}(2018)\citenamefont {Fei}, \citenamefont {Huang}, \citenamefont {Malinowski}, \citenamefont {Wang}, \citenamefont {Song}, \citenamefont {Sanchez}, \citenamefont {Yao}, \citenamefont {Xiao}, \citenamefont {Zhu}, \citenamefont {May} \emph {et~al.}}]{fei2018two}%
  \BibitemOpen
  \bibfield  {author} {\bibinfo {author} {\bibfnamefont {Z.}~\bibnamefont {Fei}}, \bibinfo {author} {\bibfnamefont {B.}~\bibnamefont {Huang}}, \bibinfo {author} {\bibfnamefont {P.}~\bibnamefont {Malinowski}}, \bibinfo {author} {\bibfnamefont {W.}~\bibnamefont {Wang}}, \bibinfo {author} {\bibfnamefont {T.}~\bibnamefont {Song}}, \bibinfo {author} {\bibfnamefont {J.}~\bibnamefont {Sanchez}}, \bibinfo {author} {\bibfnamefont {W.}~\bibnamefont {Yao}}, \bibinfo {author} {\bibfnamefont {D.}~\bibnamefont {Xiao}}, \bibinfo {author} {\bibfnamefont {X.}~\bibnamefont {Zhu}}, \bibinfo {author} {\bibfnamefont {A.~F.}\ \bibnamefont {May}}, \emph {et~al.},\ }\bibfield  {title} {\bibinfo {title} {Two-dimensional itinerant ferromagnetism in atomically thin fe3gete2},\ }\href@noop {} {\bibfield  {journal} {\bibinfo  {journal} {Nature materials}\ }\textbf {\bibinfo {volume} {17}},\ \bibinfo {pages} {778} (\bibinfo {year} {2018})}\BibitemShut {NoStop}%
\bibitem [{\citenamefont {Mao}\ \emph {et~al.}(2016)\citenamefont {Mao}, \citenamefont {Shen}, \citenamefont {Yang}, \citenamefont {Fang}, \citenamefont {Yuan}, \citenamefont {Zhang},\ and\ \citenamefont {Liu}}]{mao2016angle}%
  \BibitemOpen
  \bibfield  {author} {\bibinfo {author} {\bibfnamefont {K.}~\bibnamefont {Mao}}, \bibinfo {author} {\bibfnamefont {W.}~\bibnamefont {Shen}}, \bibinfo {author} {\bibfnamefont {C.}~\bibnamefont {Yang}}, \bibinfo {author} {\bibfnamefont {X.}~\bibnamefont {Fang}}, \bibinfo {author} {\bibfnamefont {W.}~\bibnamefont {Yuan}}, \bibinfo {author} {\bibfnamefont {Y.}~\bibnamefont {Zhang}},\ and\ \bibinfo {author} {\bibfnamefont {X.}~\bibnamefont {Liu}},\ }\bibfield  {title} {\bibinfo {title} {Angle insensitive color filters in transmission covering the visible region},\ }\href@noop {} {\bibfield  {journal} {\bibinfo  {journal} {Scientific reports}\ }\textbf {\bibinfo {volume} {6}},\ \bibinfo {pages} {19289} (\bibinfo {year} {2016})}\BibitemShut {NoStop}%
\end{thebibliography}%

\end{document}